# Mapping the impact of papers on various status groups:

# A new excellence mapping tool based on citation and reader scores

Lutz Bornmann*, Rüdiger Mutz**, Robin Haunschild***, Felix de Moya-Anegon****,

Mirko de Almeida Madeira Clemente*****, Moritz Stefaner******

*Corresponding author:
Administrative Headquarters of the Max Planck Society
Science Policy and Strategy Department
Hofgartenstr. 8,
80539 Munich, Germany.
E-mail: bornmann@gv.mpg.de

**Center for Higher Education and Science Studies, CHESS
University of Zurich
Andreasstrasse 15,
8050 Zurich, Switzerland.

***Max Planck Institute for Solid State Research
Heisenbergstraße 1,
70569 Stuttgart, Germany

****CSIC/CCHS/IPP, SCImago Group
Communication and Information Science Faculty
University of Granada, Granada, Spain.

***** Data Visualization & Interface Design
Fritz-Reuter-Straße 8
01097 Dresden, Germany

******Truth & Beauty
Eickedorfer Damm 35
28865 Lilienthal, Germany

**Abstract**

In over five years, Bornmann, Stefaner, de Moya Anegon, and Mutz (2014) and Bornmann, Stefaner, de Moya Anegón, and Mutz (2014, 2015) have published several releases of the www.excellencemapping.net tool revealing (clusters of) excellent institutions worldwide based on citation data. With the new release, a completely revised tool has been published. It is not only based on citation data (bibliometrics), but also Mendeley data (altmetrics). Thus, the institutional impact measurement of the tool has been expanded by focusing on additional status groups besides researchers such as students and librarians. Furthermore, the visualization of the data has been completely updated by improving the operability for the user and including new features such as institutional profile pages. In this paper, we describe the datasets for the current excellencemapping.net tool and the indicators applied. Furthermore, the underlying statistics for the tool and the use of the web application are explained.

# 1 Introduction

Citation analyses are frequently used to measure the performance of research-focused institutions and universities worldwide. These analyses are undertaken in various research evaluation processes to compare selected institutions or universities (Bornmann, Bowman, et al., 2014). For example, the Leiden Ranking (see www.leidenranking.com) is a popular university ranking covering the performance of over 1000 universities worldwide (Waltman, Calero-Medina, Kosten, Noyons, Tijssen, et al., 2012). The ranking is exclusively based on bibliometric data. The SCImago Institutions Ranking (SIR, see https://www.scimagoir.com) is another example. Some years ago, a group of researchers and experts in data visualization started to develop mapping tools to present institutional bibliometric data not only as ranking lists, but also spatially on worldwide maps. These maps (see www.excellencemapping.net) were intended to identify areas with above or below average institutional performance (Bornmann, Stefaner, de Moya Anegon, et al., 2014; Bornmann, Stefaner, de Moya Anegón, et al., 2014; Bornmann et al., 2015). Another objective of the group's activities was to spatially visualize collaborations between authors from different institutions and to reveal whether or not institutions profit from collaborations with other institutions. The collaboration map can be found at www.excellence-networks.net (see Bornmann, Stefaner, de Moya Anegón, & Mutz, 2016). Since a few years ago, university rankings, such as the Leiden Ranking and SIR, have also started to visualize the performance data spatially and to complement ranking lists.

This paper is intended to present the new and completely revised release of www.excellencemapping.net. We have extended the tool by measuring impact not only on citing authors, but on reading students, librarians, and professors. Since governments and funding organizations are increasingly interested in demonstrating the societal impact of research activities besides scientific excellence (Hicks, Stahmer, & Smith, 2018), companies

– e.g., Altmetric.com (McLeish, 2021) or Plum Analytics (Gorraiz & Gumpenberger, 2021) – have started to collect meta-data of online activities around scientific publications (Costas, 2017). Today, the collected and analyzed meta-data are known as alternative metrics (altmetrics) (Bornmann, 2014). "These alternative metrics include web citations in digitised scholarly documents (e.g. eprints, books, science blogs or clinical guidelines) and, more recently, altmetrics derived from social media (e.g. social bookmarks, comments, ratings and tweets)" (Wilsdon et al., 2015, p. 13).

Altmetrics has become so popular in recent years that conferences are organized around altmetrics research (see www.altmetricsconference.com) and the new journal *Journal of Altmetrics* (see www.journalofaltmetrics.org) has been founded. González-Valiente, Pacheco-Mendoza, and Arencibia-Jorge (2016) regard altmetrics research as being in the transition process to becoming a separate sub-discipline within the scientometrics domain. Sugimoto, Work, Larivière, and Haustein (2017) published a comprehensive literature overview of research on altmetrics. According to Blümel, Gauch, and Beng (2017), there are two dominant lines of this research: "the first kind of topics are 'coverage studies' of articles with mentions in social media platforms and their intensity … The second type of studies is cross validation studies that employ comparisons of altmetric data sources with traditional measures of scholarly performance such as citations". The meta-analysis of Bornmann (2015) summarized these cross validation studies.[1] Today, altmetrics covers a broad range of different data sources which are processed in various indicators. Moed (2017) classifies these sources as follows:

- "Social media such as Twitter and Facebook, covering social activity.
- Reference managers or reader libraries such as Mendeley or ResearchGate covering scholarly activity.

---

[1] The results are as follows: "the correlation with traditional citations for micro-blogging counts is negligible (pooled $r = 0.003$), for blog counts it is small (pooled $r = 0.12$) and for bookmark counts from online reference managers, medium to large (CiteULike pooled $r = 0.23$; Mendeley pooled $r = 0.51$)" (p. 1123).

- Various forms of scholarly blogs reflecting scholarly commentary.
- Mass media coverage, for instance, daily newspapers or news broadcasting services, informing the general public" (p. 68).

One of the most controversially discussed and empirically studied questions around altmetrics is the kind of impact being measured. Are altmetrics able to measure societal impact of research (see Barthel, Tönnies, Köhncke, Siehndel, & Balke, 2015; Blümel et al., 2017)? Bornmann, Haunschild, and Adams (2019) compared altmetrics with societal impact assessments of research by peers and suggested that altmetrics "may capture a different aspect of societal impact (which can be called unknown attention) to that seen by reviewers (who are interested in the causal link between research and action in society)" (p. 325). Their assessment corresponds to the assessments by other authors: "The current author agrees with the proposition that usage-based indicators and altmetrics primarily reflect attention rather than influence" (Moed, 2017, p. 133). Kassab, Bornmann, and Haunschild (2020) casted doubt on the usability of specific altmetrics (counts of mentions in Twitter, Facebook, blogs, news outlets, policy documents, and Wikipedia) for measurement of societal impact. The results of Haunschild, Leydesdorff, Bornmann, Hellsten, and Marx (2019) show that "Twitter networks seem to be able to visualize public discussions about specific topics" (p. 695). According to Konkiel, Madjarevic, and Rees (2016) "altmetrics can tell us about public influence and non-traditional scholarly influence" (p. 16).

Mendeley is a web-based reference manager (free-to-use) which can be used to save the bibliographic information of publications with the intention of reading and citing them or using them for other purposes, e.g., for teaching (Haunschild, 2021). Mendeley is a popular altmetrics source; Mendeley counts measure the number of times publications have been bookmarked in Mendeley libraries. Section 2.1 provides an overview of scientometric studies based on Mendeley data (the data have been used already in many previous empirical studies). The studies – as has been pointed out by Zahedi and Haustein (2018) – reveal that "Mendeley readership and citations capture a similar concept of impact, although they cannot be considered as equivalent indicators" (p. 191). Research has shown that Mendeley has the "most extensive coverage of scientific literature" (Pooladian & Borrego, 2016, p. 1136) compared to other altmetrics. Thus, many papers that can be found in multidisciplinary databases, such as Web of Science (WoS, Clarivate Analytics) or Scopus (Elsevier), are also covered in Mendeley. The good coverage of the literature in Mendeley induced Bornmann and Haunschild (2016) and Haunschild and Bornmann (2016) to develop field-normalized readership scores – following the standard field-normalizing approaches used in bibliometrics (Waltman, 2016) – for using Mendeley impact data for cross-field-comparisons.

A further advantage of Mendeley readership data besides coverage is that some information on the Mendeley users is available who saved the publications in their libraries (Gunn, 2013), such as nationality, status and discipline (Maflahi & Thelwall, 2018). The most interesting information is the status of the users (e.g., professor, student, or librarian) which encouraged Bornmann and Haunschild (2017) to develop field-normalized Mendeley indicators measuring impact of research on specific user groups. Since these indicators significantly extend the broadness of impact measurements, we decided to include these indicators in the new release of www.excellencemapping.net. Thus, the new release is able to visualize the impact of institutional research not only in the traditional way (by using citation counts), but also in the new way by demonstrating institutional impact targeted to specific groups of the society.

In the following sections, we describe the literature in the area of using Mendeley readership data for research evaluation and the use of bibliometric data for spatial visualizations. Then, the datasets for the current excellencemapping.net tool and the indicators applied are described. The underlying statistics for the tool and the use of the web application are explained in the results section.

# 2 Literature overview

The literature overview has two parts that focus on two relevant topics for this study: (1) research on Mendeley data being used as an altmetrics source, and (2) studies that spatially map scientometrics data.

## 2.1 Research on Mendeley

Mendeley is a free web-based reference manager that has been available since 2007. Users can enter information about publications of interest (e.g., for later reading, sharing, or citing; see Thelwall & Kousha, 2015). According to the literature overview of Thelwall and Wilson (2016), Mendeley users may be biased towards younger people (students or PhD students). The authors further claim that "it is reasonable to think of Mendeley users who bookmark an article in the site as readers of the article because most of them have read, or intend to read, articles that they bookmark" (Thelwall & Wilson, 2016, p. 1964). The survey results by Mohammadi, Thelwall, and Kousha (2016) might confirm this claim (see Thelwall & Kousha, 2015). Based on this and similar results, Mendeley data have been frequently denoted as readership data. Most important (for the use of Mendeley data in the current study) is the result by Mohammadi et al. (2016) that "about 85% of the respondents across all disciplines bookmarked papers in Mendeley to cite them in their publications" (p. 1202). Thus, it seems that Mendeley data are a "byproduct of existing workflows" (Haustein, 2014, p. 339), which emerges as an element of the scholarly communication process.

As already mentioned in section 1, research on Mendeley mostly addressed two topics: coverage of the literature in Mendeley and correlation of readership counts with citation counts.

2.1.1 Coverage of the literature

In their review of research on altmetrics, Sugimoto et al. (2017) included an overview of studies investigating the coverage of Mendeley. Since this overview is based on many previous studies, the results can be assumed to be still valid. The different studies reported coverage shares (share of papers having at least one reader) as being between 60% and 80%. Factors that lead to differences in these shares are the investigated journal, field, and data aggregator. Coverage seems to be higher for more recent publications and papers published in high impact journals (e.g., *Science* or *Nature*). In a comprehensive recent study, Zahedi and van Eck (2018) investigated the coverage not only by field, but also by Mendeley status group. They found that "publications from Mathematics & Computer Science have the lowest coverage in Mendeley. Publications from the Social Sciences & Humanities receive on average the highest number of readers in Mendeley". With respect to coverage among different status groups, the highest coverage was found for students (87.9%). "This is followed by researchers (70.3%) and professors (63.6%). The lowest coverage is for other professionals (33.2%) and librarians (10.0%)" (Zahedi & van Eck, 2018).

2.1.2 Correlation with citation counts

The meta-analysis published by Bornmann (2015) covered (many) previous empirical studies correlating Mendeley scores and citation counts. The pooled $r = 0.51$ reveals a substantial correlation between both metrics. Later studies confirmed this extent of relationship between both metrics. For example, Pooladian and Borrego (2016) report nearly the same correlation coefficients ($r = 0.52$ for 2015 and $r = 0.56$ for 2016).

Thelwall (2017) correlated Mendeley reader scores with Scopus citations for papers published in 2012 in various subject categories. He found that "despite strong positive correlations in most fields, averaging 0.671, the correlations in some fields are as weak as 0.255" (Thelwall, 2017, p. 1721). According to the results by Aduku, Thelwall, and Kousha (2016) the correlations vary not only by field, but also by document type. Later results published by Thelwall (2018) reveal that "there are moderate or strong correlations in eight out of ten fields ... The correlations are higher than the correlations between later citations and early citations, showing that Mendeley reader counts are more useful early impact indicators than citation counts" (p. 1231). The author interpreted his findings as signals that Mendeley reader scores might be useful indicators for later citation impact and recommended Mendeley reader counts "as early impact indicators for situations where citation counts are valued as impact indicators in the fields analysed" (Thelwall, 2018, p. 1238).

In another study published in the same year, Didegah and Thelwall (2018) came to very similar conclusions: "the moderate differences between patterns of saving and citation suggest that Mendeley can be used for some types of impact assessments, but sensitivity is needed for underlying differences" (Didegah & Thelwall, 2018, p. 959). After a short literature overview of Mendeley studies, Pooladian and Borrego (2016) reach similar conclusions. The application of Mendeley reader counts as an early impact indicator might be especially useful in those situations in which an analysis is made of the impact of papers that have been published in journals with large publication delays (significant differences between the date of online availability and publication in a journal issue) (Maflahi & Thelwall, 2018). The reason is that Mendeley users add the papers to their library very early when they become aware of a paper. This point in time might be significantly earlier than the point in time when the users read or cite the paper (Thelwall, 2017).

Although altmetrics aggregators (such as Altmetric.com or Plum Analytics) offer Mendeley data, they should be retrieved directly from Mendeley because Mendeley data from altmetrics aggregators are often incomplete (Bar-Ilan, Halevi, & Milojević, 2019).

2.1.3   Possible disadvantages of using Mendeley data as altmetrics data

Although Mendeley seems to be one of the most interesting altmetrics sources, it is not without limitations, which should be considered in the interpretation of empirical results based on Mendeley data. The most important limitations of this altmetrics source are listed in Mas-Bleda and Thelwall (2016) and Thelwall (2017):

(1) It is possible to game Mendeley reader counts (without greater expense). In contrast to citations where manuscripts have to be published to push the citations of certain publications, for reader counts, it is simply necessary to include certain publications – which should be pushed – in user libraries. Thus, the hurdle for gaming is significantly lower for Mendeley reader counts than for citations. For example, "Mendeley readership counts could be deliberately spammed by publishers or authors creating many artificial Mendeley accounts to bookmark a set of articles" (Thelwall & Wilson, 2016, pp. 1969-1970).

(2) The Mendeley reference manager is only one tool in addition to certain other tools that are available (e.g., Endnote or BibTeX). Thus, if only Mendeley data are used for research assessments, literature which is stored in libraries of other tools is not considered (van Noorden, 2014). However, no studies are available (to the best of our knowledge) which showed that the focus on one tool produces a systematic bias. A similar problem concerns the fact that not all academics use a reference manager. Some academics refrain from using a software solution or have designed their own solution.

(3) As already mentioned above, Mendeley users are biased towards younger academics (PhD students, postgraduates, and postdocs). Mendeley also has an international bias, since the uptake of the tool differs by nations (Haunschild, Stefaner, & Bornmann, 2015). Furthermore, it appears that Mendeley users tend to save publications from their own countries in libraries (Maflahi & Thelwall, 2015).

According to Haustein (2014), another problem with Mendeley data – not mentioned by Mas-Bleda and Thelwall (2016) and Thelwall (2017) – is the quality of the data: "A more serious problem is the incompleteness and errors that are found in the metadata of bibliographic entries in online reference managers. This often causes an article bookmarked by more than one user not to be recognized as one and the same publication" (p. 336).

## 2.2 Studies on spatial bibliometrics

Science maps (also known as scientographs, bibliometric network visualizations, or knowledge domain maps) are "visual representations of the structure and dynamics of scholarly knowledge … Science maps are usually generated based on the analysis of large collections of scientific documents" (Petrovich, 2020). Thus, these maps cannot be used to analyze research that is practiced outside the publication domain. Science maps are applied in various disciplines, such as information science and the sociology of science. Science mapping "stands at the crossroad of numerous disciplines: scientometrics, library and information science, citation analysis, text analysis, statistics, network analysis, among others" (Petrovich, 2020).

Frenken, Hardeman, and Hoekman (2009) introduced the term "spatial scientometrics" – one branch in the area of science mapping – which is defined as a "combination of the domain analysis method originated in information science and visual analytics from computer science" (Chen & Song, 2017, p. 106). Frenken et al. (2009) published a comprehensive first review of studies in spatial scientometrics. These authors explained the research area of spatial scientometrics as "quantitative science studies that explicitly address spatial aspects of scientific research activities" (p. 222). Five years later, two of the three previous authors published a second review covering the literature from recent years (Frenken & Hoekman, 2014). Frenken and Hoekman (2014) concluded in that second review that "the field of scientometrics has witnessed a rapid increase in studies using spatial data" (p. 126). Our own observation is that this tendency has been reinforced since 2014. As we have already pointed out in section 1, available university rankings have started to routinely present a spatial visualization of performance data.

Frenken and Hoekman (2014) assume that the interest in spatial bibliometrics is mainly triggered by two debates in research evaluation: (1) emerging economies such as India, Brazil, and especially China have become important players in science, and science managers and researchers are interested in their development within global national comparisons. For example, Aumüller and Rahm (2011) show the geographic distribution of the main institutions contributing research results in computer science. (2) The second debate concerns the so-called "European Paradox": "For a long time it was assumed that European countries were global leaders in terms of impact as measured by citation counts, but lagged behind in converting this strength into innovation, economic growth and employment" (Frenken & Hoekman, 2014, p. 131). Studies (e.g., Rodríguez-Navarro & Narin, 2018) have used various datasets to test whether this paradox really exists.

Besides studies visualizing institutional data, there is an increasing interest in spatial distributions of performance data at the level of cities and sub-national regions. According to Frenken and Hoekman (2014), this is likely to be due to the fact "that larger data efforts are required to accurately classify addresses from scientific papers into urban or regional categories as well as from the fact that science policy is mainly organized at national and transnational levels" (p. 132). For example, Csomos (2018) used Scopus data to examine and visualize scientific output in the period from 1986 to 2015 at the city level. The analysis confirms the rapid increase of publication output for many Chinese cities. In another study, Grossetti et al. (2013) investigated whether scientific activities in countries are more or less strongly focused on large cities. The authors found that "deconcentration is the dominant trend both globally and within countries" (p. 2219). In their paper, Grossetti et al. (2013) extensively discussed the problems of defining city areas for the purpose of spatially mapping performance data. As an example of spatial bibliometrics on the sub-national level, Hu, Guo, and Hou (2017) analyzed the performance of 31 Chinese regions. They identified regions with high publication output and research preferences of the regions.

Other studies did not focus on the performance of entities (institutions or cities), but rather the cooperation between different entities. The overview by Maisonobe, Eckert, Grossetti, Jégou, and Milard (2016) identifies three main objectives of these studies: "the reasons for scientific collaboration, the proximity and size effects on the propensity to collaborate, and longitudinal tendencies of the world collaboration network" (p. 1026). Maisonobe et al. (2016) used publication data from 2000 to 2007 to study the development of inter-urban collaborations. They found a declining trend of publications produced in single cities, and an increase in the number of publications produced by authors from different cities. However, these authors were located tendentially in the same country (and thus are not international collaborations). Some years later, Maisonobe, Jégou, Yakimovich, and Cabanac (2019) published the NETSCITY tool that can be used to map global-scale production and collaboration data between cities. In a very recent study, Csomos and Lengyel (in press) investigated the efficiency of scientific collaborations between cities in the period between 2014 and 2016. The results show that "US-Europe co-publication links are more efficient in terms of producing highly cited papers than those international links that Asian cities have built in scientific collaboration".

## 3   Methods

### 3.1   Bibliometric data and indicators

We used the publication and citation data from the SIR. This tool has been developed by the SCImago Lab for bibliometric analysis with the aim of measuring the performance of universities and research-focused institutions worldwide. The data processing includes the extensive semi-automatic and manual identification and disambiguation of all universities and research-focused institutions based on the institutional affiliation of the publications that are included in the Scopus database (Elsevier). Further extensive information on the SIR data processing can be found at www.scimagoir.com/methodology.php.

For the excellence mapping tool, we use the excellence indicator from SIR, which is a field- and time-normalized percentile indicator (Bornmann, de Moya Anegón, & Leydesdorff, 2012; Bornmann & Williams, 2020). In order to calculate the excellence rate, SIR orders the publications in a subject area (ASJC, all science journal classification – the Scopus journal classification system)[2] and publication year combination by (1) citation counts and (2) the SCImago Journal Ranking (SJR, www.scimagojr.com/journalrank.php) citation impact value of the journal in which the publication appeared (Gonzalez-Pereira, Guerrero-Bote, & Moya-Anegon, 2010). The results by Bornmann, Leydesdorff, and Wang (2013) show that the excellence rate has "higher power in predicting the long-term citation impact on the basis of citation rates in early years" (p. 933) than other field-normalized approaches. To identify highly cited papers, the SJR value is used as a second criterion besides citation counts in order to resolve ties in citation counts. In each combination of subject area and publication year, the 10% most frequently cited publications are selected as highly cited publications. In a small number of certain combinations, not exactly 10% are selected, since ties in citation data could not be completely resolved in every case.

Since we consider in this study also top 10% indicators which are based on Mendeley data, we do not use the SIR name "excellence indicator" in this study, but PP(top 10%) for all these top 10% indicators.

**3.2  Mendeley data and indicators**

Out of all papers (n=10,260,559) published between 2012 and 2016, 8,928,486 unique DOIs occurred. For 15,009 papers, non-unique DOIs were observed. We removed these papers with duplicated DOIs from the set of DOIs to query the Mendeley API. The Mendeley reader counts and their academic status group information (where available) were downloaded via the Mendeley API using R (R Core Team, 2014). We used the R packages httr (Wickham, 2017a), rjson (Couture-Beil, 2014), RCurl (Lang & the CRAN team, 2018), and yaml (Stephens et al., 2018) for downloading the Mendeley data. The Mendeley data were requested between 20 October 2020 and 25 October 2020 for 8,913,477 DOIs from the Mendeley API. For 8,912,090 DOIs, we received valid Mendeley user data, and for 1,387

---

[2] see https://service.elsevier.com/app/answers/detail/a_id/15181/supporthub/scopus/

DOIs, an error occurred. On 25 October 2020, we attempted to retrieve the Mendeley data for these DOIs in a second round. Of these 1,387 DOIs, we found Mendeley data for 102 DOIs in this second run. We attempted to retrieve Mendeley data for the remaining 1,285 DOIs on the same day, but obtained an error for all of them. The retrieved Mendeley data from the first and second run were merged.

In total, we found 8,912,192 of the DOIs (99.99%) in the Mendeley API. Of those DOIs, 437,085 were registered without a Mendeley reader. We found 280,924,726 reader counts for all queried papers (on average 31.5 reader counts per paper). Mendeley provides different academic status groups for user classification: "Lecturer", "Lecturer > Senior Lecturer", "Librarian", "Professor", "Professor > Associate Professor", "Researcher", "Student > Bachelor", "Student > Doctoral Student", "Student > Master", "Student > Ph. D. Student", "Student > Postgraduate", "Unspecified", and "Other". For 248,778,788 reader counts (88.6%), academic status group information was available. We dropped the two academic status groups "Unspecified" and "Other" for the definition of Mendeley sectors. However, they do contribute to the sector "Total" because this sector contains all Mendeley reader counts, irrespective of their academic status group.

We defined the following sectors besides the sector "Total" (based on classifications proposed by Haustein & Larivière, 2014; Pooladian & Borrego, 2016):

− Lecturers: "Lecturer" and "Lecturer > Senior Lecturer" (5,847,013 reader counts)

− Librarians: "Librarian" (1,683,242 reader counts)

− Professors: "Professor" and "Professor > Associate Professor" (18,405,980 reader counts)

− Researchers: "Researcher" (35,469,732 reader counts)

- Students: "Student > Bachelor", "Student > Doctoral Student", "Student > Master", "Student > Ph. D. Student", and "Student > Postgraduate" (129,808,299 reader counts)

The R packages "data.table" (Dowle & Srinivasan, 2019), "plyr" (Wickham, 2011), and "tidyverse" (Wickham, 2017b) were used for data analysis.

We applied another procedure to calculate P(top 10%) based on Mendeley data than on citation data (see section 3.1). One reason was that it does not make sense to use the SJR value as criterion to select the 10% most frequently bookmarked papers (with cited papers it may make sense, as the results by Bornmann, Leydesdorff, et al., 2013, show). The Mendeley P(top 10%) is based on the information as to whether or not a paper has been bookmarked in Mendeley by a certain user (e.g., a professor). Thus, a zero value means that a paper has not been bookmarked by a Mendeley user. Based on the Mendeley user data, the formula derived by Hazen (1914) (($i-0.5$)/$n$*100) was used for calculating percentiles separately for each publication year and subject area (ASJC) combination. For papers without valid DOI and papers that could not be retrieved from the Mendeley API, we assigned a randomly chosen percentile value. The random percentile values were chosen individually for each ASJC in the case of papers with multiple field classifications.

In total, random percentile values were chosen for 1,348,367 papers. Those papers include 1,332,073 (98.8%) without a DOI in the data set. Thus, the random percentiles were mainly used for papers without a DOI because it is unfeasible to obtain Mendeley reader counts via the API for papers without DOIs with such a large data set. The Hazen percentiles were used to determine the 10% most frequently bookmarked papers for each of the Mendeley sectors. Following the approach proposed by Waltman and Schreiber (2013), tied papers with the same number of bookmarks at the top 10% threshold (in a certain subject area) were fractionally assigned to the 10% most frequently bookmarked papers. In order to determine the "All subject areas" value, we used the maximum of the P(top 10%) value of the individual ASJCs, i.e., the best performing subject area value was used as the "All subject areas" value.

Table 1 shows the 90[th] percentile thresholds and average values for Mendeley reader counts of the analyzed user groups broken down by ASJC code for the publication year 2012 as an example. The data for the other years are very similar. The data in Table 1 show that the results for the Librarians Mendeley reader group should be interpreted with caution, especially in the ASJC codes with 0 or 1 as a 90[th] percentile threshold. The difference between highly and lowly cited papers is only partly given by using these thresholds. In addition, the results for the Lecturers Mendeley reader group with the ASJC code 2600 (General Mathematics) should be cautiously interpreted due to the same problem. We did not include the same table for citations, since citations are not affected by these low thresholds.

Table 1. 90th percentile thresholds and average values for Mendeley reader counts of the analyzed user groups broken down by ASJC code for the publication year 2012 as an example

| ASJC code | Librarians | | Lecturers | | Professors | | Researchers | | Students | | Total | | Number of papers |
|---|---|---|---|---|---|---|---|---|---|---|---|---|---|
| | 90th perc. | average | 90th perc. | average | 90th perc. | average | 90th perc. | average | 90th perc. | average | 90th perc. | average | |
| 1000 | 1 | 0.43 | 5 | 2.37 | 24 | 12.19 | 54 | 29.22 | 127 | 74.19 | 244 | 135.27 | 99,250 |
| 1100 | 1 | 0.24 | 3 | 1.18 | 11 | 5.15 | 27 | 12.27 | 71 | 32.77 | 126 | 59.22 | 234,012 |
| 1200 | 1 | 0.35 | 5 | 1.91 | 13 | 5.96 | 20 | 10.26 | 74 | 34.54 | 133 | 60.41 | 65,893 |
| 1300 | 1 | 0.27 | 3 | 1.15 | 12 | 5.80 | 25 | 12.44 | 67 | 33.57 | 121 | 61.71 | 527,979 |
| 1400 | 1 | 0.40 | 11 | 4.46 | 17 | 6.97 | 14 | 5.66 | 124 | 51.96 | 187 | 78.87 | 45,915 |
| 1500 | 0 | 0.10 | 2 | 0.86 | 8 | 3.79 | 15 | 6.91 | 52 | 24.20 | 87 | 41.00 | 141,107 |
| 1600 | 0 | 0.09 | 2 | 0.67 | 7 | 3.32 | 13 | 5.77 | 40 | 18.58 | 67 | 32.53 | 329,057 |
| 1700 | 0 | 0.12 | 2 | 0.64 | 5 | 2.00 | 7 | 2.88 | 30 | 13.34 | 47 | 21.40 | 339,752 |
| 1800 | 1 | 0.26 | 4 | 1.66 | 10 | 4.06 | 10 | 4.33 | 63 | 25.74 | 98 | 40.85 | 29,504 |
| 1900 | 1 | 0.18 | 2 | 0.71 | 8 | 3.50 | 18 | 8.32 | 38 | 17.26 | 74 | 34.57 | 191,233 |
| 2000 | 1 | 0.29 | 7 | 2.87 | 13 | 5.48 | 13 | 5.32 | 87 | 35.89 | 135 | 56.79 | 36,327 |
| 2100 | 0 | 0.08 | 3 | 1.02 | 7 | 3.03 | 14 | 5.89 | 53 | 22.79 | 85 | 37.81 | 71,963 |
| 2200 | 0 | 0.09 | 2 | 0.62 | 5 | 2.38 | 8 | 3.78 | 29 | 12.64 | 52 | 22.81 | 545,492 |
| 2300 | 1 | 0.27 | 4 | 1.42 | 10 | 4.81 | 25 | 11.46 | 75 | 33.63 | 129 | 59.67 | 160,743 |
| 2400 | 1 | 0.29 | 3 | 1.22 | 11 | 5.10 | 25 | 12.01 | 68 | 33.37 | 122 | 61.00 | 134,460 |
| 2500 | 0 | 0.07 | 2 | 0.56 | 6 | 2.71 | 11 | 4.77 | 35 | 15.67 | 59 | 27.19 | 343,747 |
| 2600 | 0 | 0.06 | 1 | 0.46 | 4 | 1.97 | 6 | 2.84 | 19 | 8.49 | 33 | 15.83 | 229,376 |
| 2700 | 2 | 0.48 | 3 | 1.36 | 10 | 4.57 | 18 | 8.96 | 58 | 28.35 | 106 | 53.08 | 1,112,661 |
| 2800 | 1 | 0.35 | 4 | 1.51 | 16 | 7.36 | 30 | 14.36 | 88 | 42.64 | 155 | 76.72 | 127,555 |
| 2900 | 3 | 0.99 | 4 | 1.86 | 9 | 4.04 | 16 | 7.70 | 68 | 33.13 | 116 | 57.83 | 52,923 |
| 3000 | 1 | 0.26 | 2 | 0.85 | 7 | 3.24 | 14 | 6.46 | 44 | 21.02 | 77 | 37.87 | 110,925 |
| 3100 | 1 | 0.23 | 3 | 0.93 | 16 | 5.67 | 28 | 9.47 | 28 | 13.08 | 115 | 36.14 | 605,357 |
| 3200 | 2 | 0.61 | 6 | 2.44 | 13 | 6.43 | 21 | 10.08 | 100 | 49.03 | 158 | 79.40 | 78,014 |
| 3300 | 2 | 0.62 | 6 | 2.22 | 10 | 4.20 | 14 | 5.94 | 68 | 29.83 | 110 | 49.13 | 153,309 |
| 3400 | 1 | 0.31 | 4 | 1.32 | 7 | 3.03 | 13 | 6.10 | 41 | 19.19 | 80 | 37.49 | 25,071 |
| 3500 | 2 | 0.54 | 3 | 1.30 | 8 | 3.92 | 9 | 4.13 | 60 | 29.95 | 95 | 48.61 | 16,511 |
| 3600 | 2 | 0.51 | 5 | 1.85 | 10 | 4.38 | 16 | 7.28 | 72 | 32.89 | 120 | 56.76 | 50,366 |

### 3.3 Statistical model

The choice of the statistical model for analyzing the data usually depends on the scale of the dependent variable (e.g., ordinal, continuous). In this study, the dependent variable is binary: either a paper published by an author located in an institution belongs to P(top 10%), or it does not. However, it was not the binary data that were modeled (Bernoulli distribution), but the aggregated fractional data for the institutions (Binomial distribution). Since absolute frequencies are integer values in statistical procedures, the Mendeley PP(top 10%) were rounded to be included in the analyses. The PP(top 10%) for an institution is an estimate of its probability of being among the top 10% institutions. From a statistical point of view, it is not appropriate to report only the relative frequencies of individual institutions, as is done in the Leiden ranking (Mutz & Daniel, 2015; Waltman, Calero-Medina, Kosten, Noyons, Tijsen, et al., 2012). The hierarchical structure of the data (papers at level 1 are nested within institutions at level-2) is neglected. Papers published by authors of the same institution are more homogeneous regarding their probability of being P(top 10%) than papers from different institutions. Thus, the effective sample size decreases and the standard errors increase due to the increase of homogeneity within institutions.

Even the P(top 10%) probability itself would change if the sample size were small. In the logic of "Bayes estimates", there are two estimates for the probability of an institution: (1) the overall P(top 10%) probability of papers published by all institutions (in a field), if there is no information about a focal institution, and (2) the observed PP(top 10%) of the focal institution. A Bayes estimate is nothing but the mean of both estimates weighted by the reliability (Greenland, 2000). The higher the reliability in the sense of a high number of institutional papers, the higher the relative frequency is weighted; the lower the sample size and reliability, the more the overall probability is weighted. Therefore, the lower the sample size, the more the estimated probability deviates from the relative frequency for an institution.

A multilevel model is capable of handling complex data structures with a small set of parameters (fixed effects, variance components) and allows to statistically control for sets of covariates, which might bias the mean differences among institutions (e.g., Hox, Moerbeeck, & van de Schoot, 2017; Mutz & Daniel, 2007; Snijders & Bosker, 2011). Compared to the previous versions of the excellence mapping tool, not only bibliometric but also Mendeley data for different user groups are analyzed in one single model. The multivariate data structure with seven dependent variables (six Mendeley-based indicators and one bibliometric indicator) is modelled by a univariate multilevel model, as illustrated below.

As mentioned above, in multilevel regression model (MLR) papers are assumed to be nested within institutions, whereas $j$ ($j = 1, …, N$) denotes the level-2 units or clusters ("institutions") and $i$ ($i = 1, …, n_j$) the level-1 units ("papers"). The dependent variable $y_{ji}$ is binary: 1 = paper $i$ belongs to P(top 10%), while 0 = paper $i$ does not belong to P(top 10%). To simplify the analysis, we sum P(top 10%) of each institution in order to model the relative frequency, $y_j$, or probability of P(top 10%), $p_j$, of an institution. To map the multivariate data structure, the seven dependent variables ($k = 1$ to 7) and the corresponding data are set one below the other for each institution, as shown in Table 2 (see Goldstein, 2011). Each of the seven variables is identified using dummy-coding ($d_1$ - $d_7$).

Table 2. Multivariate data matrix for an example with two institutions

| Institution | Response $y_{ji}$ | $d_1$ | $d_2$ | … | $d_7$ | $x_j$ |
|---|---|---|---|---|---|---|
| 1 | $y_{11}$ Mendeley: Lecturer | 1 | 0 | … | 0 | 10 |
| 1 | $y_{21}$ Mendeley: Librarians | 0 | 1 | … | 0 | 10 |
| 1 | … | … | … | … | … | 10 |
| 1 | $y_{71}$ Citation counts | 0 | 0 | … | 1 | 10 |
| 2 | $y_{12}$ Mendeley: Lecturer | 1 | 0 | … | 0 | 45 |
| 2 | $y_{22}$ Mendeley: Librarians | 0 | 1 | … | 0 | 45 |
| 2 | … | … | … | … | … | 45 |
| 2 | $y_{72}$ Citation counts | 0 | 0 | … | 1 | 45 |

The cluster-covariate x$_j$ is taken into account in order to statistically control the differences among institutions or the ranking of institutions for this specific covariate. If the covariate was continuous, it was z-standardized (M = 0, STD = 1). The residuals among institutions can then be interpreted as if all institutions were equal according to this covariate (see Bornmann, Mutz, & Daniel, 2013). The multivariate generalized linear mixed model for the binomial data of each subject category consists of three components (Bornmann, Mutz, Marx, Schier, & Daniel, 2011; Bornmann, Stefaner, de Moya Anegon, et al., 2014; Goldstein, 2011; Hox et al., 2017), which can be formalized as follows:

1. The probability distribution for y$_j$ is a binomial distribution, bin (n$_j$, π$_j$), where π$_j$ is the expected probability and n$_j$ is the total number of papers published by an institution.

2. A linear multilevel regression part with a latent (unobserved) predictor η$_j$ without intercept (see Goldstein, 2011, p. 162) is defined as follows:

$$\eta_j = \sum_{k=1}^{K} \beta_k d_{kj} + \sum_{k=1}^{K} u_{kj} d_{kj} \qquad u_{kj} \sim N(\mathbf{0}, \mathbf{S}), \qquad (1)$$

where $\beta_k$ is the $k^{th}$ regression coefficient for $k^{th}$ dummy variable d$_{kj}$. The random effects for each institution and dummy variable are distributed according to a multivariate normal distribution, with a *k*-dimensional zero mean vector **0** and a *K*×*K* covariance matrix **S**. The matrix represents the variances and covariances of the Mendeley and bibliometric variables. For **S**, a *heterogeneous compound-symmetry structure* (CSH) was assumed with a variance component for each variable (Mendeley and bibliometric indicators) and an overall correlation coefficient for all bivariate correlations among the seven variables (see SAS Institute Inc., 2014, p. 3165).

In the case of a covariate, an interaction term is included in Eq (1):

$$\eta_j = \sum_{k=1}^{K} \beta_k d_{kj} + \sum_{k=1}^{K} \beta_{K+k} d_{kj} x_j + \sum_{k=1}^{K} u_{kj} d_{kj} \qquad u_{kj} \sim N(\mathbf{0}, \mathbf{S}), \qquad (2)$$

In order to estimate the overall mean probability of P(top 10%) for an institution across all Mendeley groups, an intercept model was additionally estimated using an overall intercept and effect coding of the six Mendeley variables.

3. The expected value of the dependent variable y is linked with the latent predictor η by the logit function: $\eta_j = \text{logit}(\pi_j) = \log(\pi_j /(1- \pi_j))$. Probabilities ranging between 0 and 1 are transformed by the logit link function to logits. They continuously vary between -∞ and +∞ with a variance of $\pi^2/3 = 3.29$.

For the visualization, the expected values of the institutional P(top 10%) probability were used instead of the raw relative frequencies. The probabilities provide empirical Bayes estimates.

The expected value of an institution $p_j$ can be estimated as follows:

$$\hat{p}_j = \text{logistic}(\sum_{k=1}^{K} \beta_k d_{jk} + \sum_{k=1}^{K} u_{jk} d_{jk}) \qquad u_{jk} \sim N(\mathbf{0}, \mathbf{S}), \qquad (3)$$

where "logistic" denotes the logistic transformation of $p_j$ (logistic(z) = $e^z/(1+e^z)$) – the inverse logit link function. In the case of a covariate, Eq (3) represents the residuals, i.e., the probabilities that are not influenced by the covariate (i.e., statistically controlled by the covariate).

An intra-class correlation for each variable, $\rho_k = \sigma^2_{uk}/(3.29+\sigma^2_{uk})$ can be calculated, which reflects the homogeneity of papers within an institution. The Wald test was used to test the null hypothesis whether $\sigma^2_{uk}$ deviates from 0. In the case of a statistically significant Wald test, the differences among institutions with respect to their PP(top 10%) are not random. In other words, it makes sense to rank institutions with respect to the selected subject category and covariate.

The proportion of variance $R^2_{uk.x}$ for each variable explained by the cluster-covariates is defined as the difference between: (1) the variance component $\sigma^2_{uk}$, if no covariate is included in the model, and (2) the variance component $\sigma^2_{uk|x}$, if the covariate x is included – divided by $\sigma^2_{uk}$: $R^2_{uk.x} = (\sigma^2_{uk} - \sigma^2_{uk|x})/\sigma^2_{uk}$.

As mentioned above, multilevel models provide so-called Empirical Bayes (EB) or shrinkage estimates that are even more accurate than their empirical counterparts, the relative frequencies (Bornmann et al., 2015; SAS Institute Inc., 2014). "The EB and the confidence intervals can be transformed back to probabilities to facilitate the interpretation of the results. The multiplication of standard errors by 1.39 instead of 1.96 results in so-called Goldstein-adjusted confidence intervals … with the property that if the confidence intervals of two institutions do not overlap, they differ statistically significantly ($\alpha = 5\%$) in their estimates, i.e., P(top 10%) probabilities. If the 95%-confidence interval does not include the mean proportion of P(top 10%) across all institutions, the authors located at this institution have published statistically significantly more or less P(top 10%) than the average across all institutions. In case of Goldstein-adjusted confidence intervals this test can only be done on the 16.3% probability level, rather than on the usual 5% level (Bornmann, Stefaner, de Moya Anegón, et al., 2014).

For each of the 24 Scopus subject categories and the category "All subject areas", a multilevel analysis was performed for each covariate with the SAS-procedure "proc glimmix" (SAS Institute Inc., 2014, p. 3049f). The model parameters were estimated by "maximum likelihood with an adaptive Gauss-Hermite quadrature". In the case of estimation problems, the SAS Macro program automatically reruns the analysis and applies a Pseudo-Likelihood estimation (PL): for example, the Residual PL (RSPL), where the expansions of the Taylor series is the vector of random effects, or the MSPL, where the expansion of the Taylor series is the mean of the random effects. In the case of large sample sizes, the differences among estimation methods vanish.

**3.4     Visualizisation**

The new release of our tool constitutes a complete rewrite, with new functionalities (see section 4.2), current JavaScript libraries, and React (see https://reactjs.org) at its core. As in previous versions, the world map and the ranked list play a central role visualizing the performance results of the regression models. We also refined the techniques for plotting our data and enhanced the interplay between both components, so that a user interaction in one is clearer reflected in the other.

The map was implemented with Mapbox GL JS (see https://www.mapbox.com/mapbox-gljs), using custom vector tiles. This improves the user experience and makes dynamic styling possible, which we used to differentiate between countries with and without data points. The vector tiles were generated based on the public domain map dataset Natural Earth (see https://www.naturalearthdata.com), which we adjusted with QGIS (see https://www.qgis.org) to correct the label placement of dependent countries. After converting the shapefiles to GeoJSON feature collections with GDAL ogr2ogr (see https://gdal.org/programs/ogr2ogr), we used the command-line utility Tippecanoe (see https://github.com/mapbox/tippecanoe) to build our vector tilesets. As part of this processing, we reduced the places to only those within an area of 25 km around the institutions. A shaded relief, which we obtained from the Natural Earth project, adds subtle geographical features to the map.

In the ranked list of the tool, we replaced the strips, which visualized the indicator values, with dots. This change helped us to include a series of values for multiple status groups through scattering and allows the elements to be interactive. We further improved this plot by integrating an institution's number of published papers, which is encoded by the size of each dot. This provides the user with new means to explore patterns in productivity and performance and their relationship.

In the graphical representations of the tool, color encodes an institution's performance. The continuous, diverging color scale was constructed by selecting two complementary hues for values below (red) and above (blue) the mean estimated value. For a visually balanced result, both colors have equal chroma (representing the colorfulness or intensity of the color), and luminance values in the perceptually based HCL color space. A helpful tool for this task is the hclwizard (see http://hclwizard.org). In order to make proportional differences relative to the mean in either direction perceptually equidistant, we apply a logarithmic scale to translate the values to our color space. For color interpolation, the D3.js CIE Lch (ab) interpolator (see https://github.com/d3/d3-interpolate) is employed.

### 3.5 Criteria for the selection of institutions

In view of the large number of universities and research-focused institutions worldwide, we decided not to present all institutions in the new release of www.excellencemapping.net. For example, some institutions published only a few papers in the publication years under consideration. Performance measurements based on only a few papers lead to unreliable results. We selected institutions using the following criteria:

- At least 50 institutions should be available for a subject area to be presented.
- An institution should have published at least 500 papers in a subject area.
- Institutions are presented on the highest aggregation level. For example, the Max Planck Society is included as organization and not as single Max Planck institutes.
- An institutions should have published in at least five subject areas to be included in the category "All subject areas". We use the threshold to prevent institutions that have an excellent performance in only one single subject area from outperforming institutions which are active in many subject areas.

# 4 Results

## 4.1 Results of an example model

It would go far beyond this paper to report the results of all 7×24 analyses for the excellence mapping tool (see Table 2). Instead, we report the results for the Scopus subject category "Chemistry" (ASJC = 1600) and the covariate "Gross national income per capita" (GNI). The data refer to the publication year 2017. The results of the example model are shown in Table 3. Regarding the fixed effects, there are small differences between the Mendeley and bibliometric indicators. PP(top 10%) scores vary around .11 except for the librarian Mendeley group with a lower value (.07). Since GINI is z-transformed (M = 0, STD = 1), the interactions GINI × Mendeley status group / Citation counts vanish, if the mean (=0) is chosen as value for GINI. As a result, the regression coefficients $\beta_1$ - $\beta_7$ represent the effects of each Mendeley status group / Citation counts or directly the logit-scaled proportion of each group (since an intercept is missing). A covariate (e.g., GINI) does not appear as a single covariate in the model, but appears only in the interactions.

Table 3. Estimated model parameters for "Chemistry" – I fixed effects (logit-scaled)

| Effect | Param. | Est. | SE | t-test | 95% CI | Probability, if GNI=0 |
|---|---|---|---|---|---|---|
| *I Main effects* | | | | | | |
| Mendeley group | | | | | | |
| Lecturers | $\beta_1$ | -1.90 | 0.014 | -132.28* | [-1.92; -1.86] | .13 |
| Librarians | $\beta_2$ | -2.62 | 0.220 | -119.01* | [-2.66; -2.57] | .07 |
| Professors | $\beta_3$ | -2.01 | 0.017 | -117.40* | [-2.05; -1.98] | .12 |
| Researchers | $\beta_4$ | -2.11 | 0.019 | -112.17* | [-2.14; -2.07] | .11 |
| Students | $\beta_5$ | -2.15 | 0.017 | -124.11* | [-2.19; -2.12] | .10 |
| All readers | $\beta_6$ | -2.13 | 0.017 | -123.03* | [-2.17; -2.10] | .11 |
| Bibliometrics | $\beta_7$ | -2.05 | 0.019 | -110.59* | [-2.08; -2.01] | .11 |
| *II Interactions* | | | | | | |

| | | | | | |
|---|---|---|---|---|---|
| Mendeley group | | | | | |
| GNI×Lecturers | $\beta_8$ | 0.23 | 0.014 | 16.90* | [0.21; 0.26] |
| GNI×Librarians | $\beta_9$ | 0.18 | 0.021 | 8.72* | [0.14; 0.23] |
| GNI×Professors | $\beta_{10}$ | 0.53 | 0.017 | 31.67* | [0.49; 0.56] |
| GNI×Researchers | $\beta_{11}$ | 0.70 | 0.018 | 38.58* | [0.67; 0.74] |
| GNI×Students | $\beta_{12}$ | 0.55 | 0.168 | 31.77* | [0.50; 0.57] |
| GNI×All readers | $\beta_{13}$ | 0.56 | 0.168 | 33.37* | [0.53; 0.59] |
| Bibliometrics× GNI | $\beta_{14}$ | 0.32 | 0.017 | 18.02* | [0.29; 0.36] |

Notes: Param. = parameter, Est. = estimate, SE = standard error, t-test =t-test value, 95-CI = 95% confidence interval, GNI = Gross national income per capita 2017 (z-transformed).

* p<.05

The interactions of the dummy-coded variables with the covariate are statistically significant given the null hypothesis of zero effects. There is a strong positive effect of GNI on PP(top 10%) with respect to both Mendeley and bibliometric indicators. The higher the GNI of the country (in which several institutions are located as a rule), the more the country's GNI deviates from the average GNI of all countries, and the higher are the PP(top 10 %) scores of the institutions in that country. This is especially the case for the Mendeley group of researchers ($\beta_{11}$ = 0.70), but less so for librarians ($\beta_9$ = 0.18) and lecturers ($\beta_8$ = 0.23).

This strong GNI effect is also reflected in the coefficient of determination $R^2$ (see Table 4). The coefficient measures the variance of the dependent variables that is explained by the covariate. The highest value of $R^2$ = .62 was found for the Mendeley researchers group; the lowest value with $R^2$ = .12 for librarians. About 62% (12%) of the PP(top 10%) variance of the Mendeley researchers (librarians) indicators were explained by the GNI. The variance components were still statistically significant – given the null hypothesis of a zero value and after controlling for the covariate. In other words, even if GNI is included as covariate, there is sufficient residual variability among institutions beyond chance in order to compare institutions. Furthermore, the intra-class correlations vary between .05 (lecturer) and .11 (librarians). These results reveal that the differences among the institutions are not purely random. There is sufficient institutional variability to allow inter-institutional comparisons.

Table 4. Estimated model parameters for "Chemistry" – II random effects (matrix **S**)

| Effect | Param. | Est. | SE | z-test | ICC | $R^2$ |
|---|---|---|---|---|---|---|
| *I Variance components* | | | | | | |
| Mendeley group | | | | | | |
|   Lecturers | $\sigma^2_1$ | 0.17 | 0.009 | 20.08* | 0.05 | 0.25 |
|   Librarians | $\sigma^2_2$ | 0.41 | 0.022 | 19.20* | 0.11 | 0.12 |
|   Professors | $\sigma^2_3$ | 0.25 | 0.012 | 20.45* | 0.07 | 0.52 |
|   Researchers | $\sigma^2_4$ | 0.30 | 0.015 | 20.48* | 0.08 | 0.62 |
|   Students | $\sigma^2_5$ | 0.25 | 0.012 | 20.47* | 0.07 | 0.52 |
|   All readers | $\sigma^2_6$ | 0.25 | 0.010 | 20.56* | 0.07 | 0.55 |
| Bibliometrics | $\sigma^2_7$ | 0.29 | 0.015 | 19.18* | 0.08 | 0.26 |
| *II Covariance component* | | | | | | |
| CSH correlation | $\rho$ | 0.72 | 0.011 | 64.66 | | |

Notes: Param. = parameter, Est. = estimate, SE = standard error, z-test = Wald z-test value, ICC = intra-class correlation, $R^2$ = coefficient of determination.

* p<.05

The average correlation between the seven Mendeley and bibliometric indicators amounts to .72 (see Table 4). This means that there are high inter-correlations among the variables: the higher the citation impact, the higher the reader impact on the various Mendeley status groups.

## 4.2 The web application

The URL of the web application is http://www.excellencemapping.net. Table 5 shows the number of institutions in the various subject areas that have been included in the tool after applying the selection criteria (see section 3.5). Institutions might occur multiple times in different subject areas.

Table 5. Frequency and percentage of institutions in various subject areas

| Subject area | Frequency | Percent |
|---|---|---|
| All subject areas | 1024 | 6.52 |
| Agricultural and Biological Sciences | 761 | 4.85 |
| Arts and Humanities | 260 | 1.66 |
| Biochemistry, Genetics and Molecular Biology | 1151 | 7.33 |
| Business, Management and Accounting | 128 | 0.82 |
| Chemical Engineering | 454 | 2.89 |
| Chemistry | 922 | 5.87 |
| Computer Science | 1081 | 6.88 |
| Earth and Planetary Sciences | 552 | 3.52 |
| Economics, Econometrics and Finance | 73 | 0.46 |
| Energy | 261 | 1.66 |
| Engineering | 1452 | 9.25 |
| Environmental Science | 553 | 3.52 |
| Immunology and Microbiology | 360 | 2.29 |
| Materials Science | 956 | 6.09 |
| Mathematics | 727 | 4.63 |
| Medicine | 2034 | 12.95 |
| Neuroscience | 329 | 2.10 |
| Nursing | 111 | 0.71 |
| Pharmacology, Toxicology and Pharmaceutics | 292 | 1.86 |
| Physics and Astronomy | 1218 | 7.76 |
| Psychology | 231 | 1.47 |
| Social Sciences | 616 | 3.92 |
| Veterinary | 59 | 0.38 |
| Health Professions | 96 | 0.61 |

There is a link displayed in the upper left section of the web application, "About this web application", which leads to detailed information about the tool. The page with the description includes the affiliations of the authors of the web application and a link to this research paper. The link to the description is embedded in the main toolbar of the application. This toolbar includes the following choices, from left to right: (1) Subject area: the user can select from 24 subject areas for the visualization (including "All subject areas"). (2) Measure: there is another choice between two metrics. The first metric (highly cited papers) is the probability of publishing highly cited papers [PP(top 10%)]. The second metric (highly bookmarked papers) refers to bookmarks of papers on Mendeley, and is the probability of being very frequently bookmarked (by certain status groups). For each of these measures, the tool shows the residues from the regression model (random effects) converted into probabilities. In order to have values on the original scale for both metrics (i.e., proportions of high impact papers), the intercept was added to the residues (see section 3.3). (3) Audience: If the user has selected "Highly bookmarked papers", he/she can additionally select the status group for which the results are presented (e.g., students or professors).

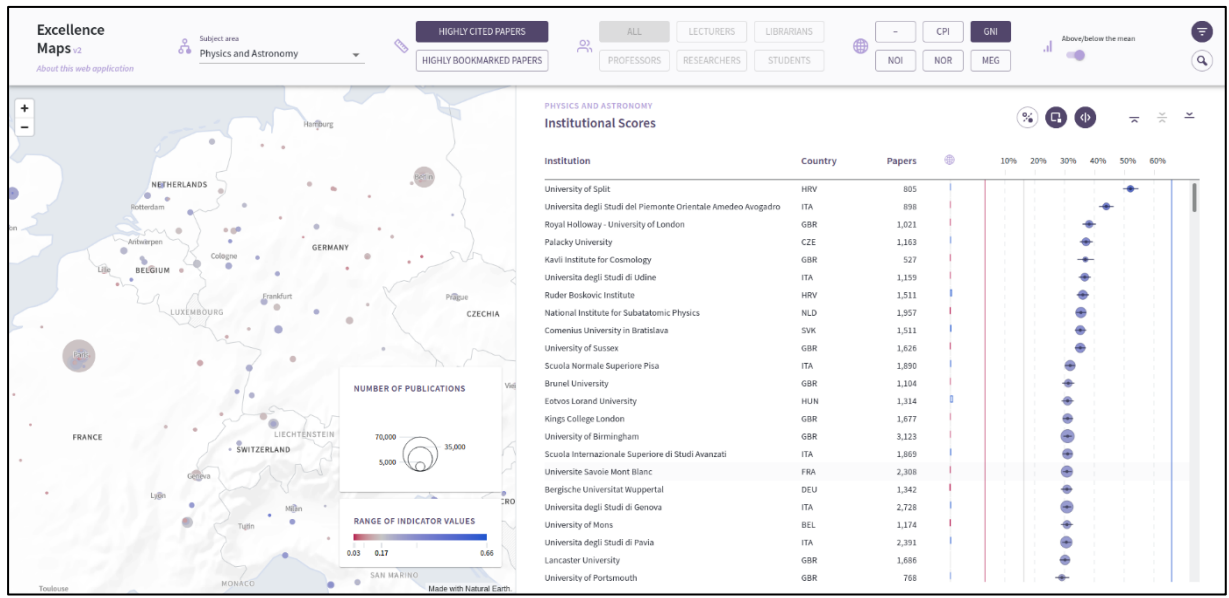

Figure 1. List and map of institutions of the excellence mapping tool

(4) Covariate: If the user selects a covariate [Number of institutions (universities or research-focused institutions) located in a country (NOI), Number of residents in a country (NOR), Gross national income per capita (GNI), Mean economic growth (MEG), Corruption perception index (CPI)], the probabilities of (i) publishing highly cited papers or (ii) publishing highly bookmarked papers (by a certain status group) is displayed adjusted (controlled) for the selected covariate. The results on the performance of institutions can then be interpreted as if the institutions all had the same value (reference point) for the covariate in question. Each covariate was z-transformed over the whole data set (with M=0 and S=1), so that the average probability shows the value in which the covariate in question has the value 0 (i.e., exactly equivalent to the median). This allows the results of the model with and without the covariates to be compared.

On the right-hand side of the main tool bar, users can tick the side bar "Above/below the mean" to reduce the set of visualized institutions in a subject area to only those which differ statistically significantly in their performance from the mean value. Here, the user also finds the possibility of searching for certain institutions included in the excellence mapping.

The performance results of the regression models for the institutions are presented in the web application on a map and in a list. The map shows a circle for each institution with a paper output greater than or equal to 500 for a selected subject area (e.g., "Physics and Astronomy"). Users can move the map to different regions with the mouse (click and drag) and zoom in (or out) with the mouse wheel. Country and city labels and map details appear only at zoom levels of a certain depth, primarily in order to facilitate perception of the data markers. Zooming can also be done with the control buttons at the top left of the map. The circle area for each institution on the map is proportional to the number of published papers in the respective subject area (or all subject areas). For example, the Centre National de la Recherche Scientifique (CNRS) has the largest circle (in Europe) on the Physics and Astronomy map, highlighting the high output of papers in this subject area.

The color of the circles on the map indicates the metric value for the respective institution using a diverging color scale, from blue through grey to red (without any reference to statistical testing): if the metric value for an institution is greater than the mean (expected) value across all institutions, its circle has a blue tint. Circles with red colors mark institutions with metric values lower than the mean. Grey circles indicate a value close to the expected value. The spectrum of the colors and corresponding metric values for a displayed map are shown in the legend of the map. The size of the circles reflects the number of papers published by an institution. Another legend of the map supports the appropriate interpretation of these sizes.

All those institutions which are taken into account in the multi-level model for a subject area (section "Institutional scores") are listed besides the map. The name, the country, and the total number of papers published ("Papers") are displayed for each institution. In addition, the probabilities of (i) publishing highly cited papers or (ii) publishing highly bookmarked papers (by a certain status group) are visualized as PP(top 10%) values between 0% and 100%. The greater the confidence interval of the probability, the more unreliable the value for an institution is. If the confidence interval does not overlap with the mean proportion across all institutions (the mean is visualized by a grey line more or less in the middle of the graph), this institution has published a statistically significantly higher (or lower) PP(top 10%) than the average across all the institutions ($\alpha = 0.165$). If the confidence intervals of two institutions do not overlap, they differ statistically significantly on the 5% level in PP(top 10%).

If the results based on Mendeley data are presented, not only the results for the selected status group are presented (with filled circles), but also the results for all other status groups and the average across all status groups (with unfilled circles). Thus, the user can compare the results for the selected status group with all other status groups (and the average across the groups).

The institutions in the list can be sorted (in descending or ascending order in the case of numbers) by clicking on the relevant heading. Thus, the top or worst performers in a subject area can be identified by clicking on the axis labels for PP(top 10%). Clicking on "Papers" puts the institutions with high productivity in terms of paper numbers at the top of the list (or at the end). In "Biochemistry, Genetics and Molecular Biology", for example, the institution with the highest productivity between 2012 and 2016 is the CNRS; in terms of highly cited papers, the best-performing institution is the Whitehead Institute for Biomedical Research. If the user selects a certain covariate, an additional column is visualized (with the symbol ): it shows for each institution by how many rank places it goes up (blue bar) or down (red bar) compared to the results based on the model without the covariate. For example, the Chinese Academy of Sciences (CAS) improves its position by 414 places compared to the ranking which does not take the covariate "Corruption perception index" into account in "All subject areas". The ranking differences in this column always relate to all institutions included. Therefore, the differences do not change if one looks at only the statistically significant results.

The symbols on the right-hand side above the axis labels have the following meaning:

Group institutions in the list by country

Focus institutions in the list to only those presented in the map segment

Adjust the scale to the listed institutional values

Scroll to the top of the listed institutions

Scroll to the bottom of the listed institutions

Scroll to the selected institution

An institution on the list or on the map can be clicked and the profile of the institution can be inspected. In the profile view, the institutional performance of the selected metric (covariate and status group in the case of Mendeley data) is presented in a graph for every subject area in which the institution has been published in the period. In the case of highly cited papers, the PP(top 10%) values are presented. In the case of highly bookmarked papers, not only the performance of the selected status group is presented, but that of all status groups in addition to an average performance across all status groups.

In addition to the graph, a map view is presented which focuses on the selected institution and institutions nearby. If the user clicks on a certain subject area in the profile view, the corresponding performance of the focal institution and all other institutions on the map (list) is presented. The number of institutions presented in the list of institutions depends on the map segment: the more institutions that are visible on the map, the more are included in the list.

Clicking on "SCImago IR" below the institution's name in the profile view opens the institutional profile view in the SIR (see www.scimagoir.com).

## 5  Discussion

The visualization of bibliometric data has become increasingly popular in recent years. For example, the Research on Research Institute (RoRI) published web-based tools that interactively visualized research landscapes based on Dimensions data (Waltman, Rafols, van Eck, & Yegros, 2019). Salinas, Giorgi , Ponchio, and Cignoni (2019) and Majeti et al. (2020) published tools for the visualization of performance data of single scientists. In a period of over five years, Bornmann, Stefaner, de Moya Anegon, et al. (2014) and Bornmann, Stefaner, de Moya Anegón, et al. (2014); Bornmann et al. (2015) have published the excellencemapping.net tool revealing (clusters of) excellent institutions worldwide based on citation data. With the new release, a completely revised tool has been published, which is not only based on citation data (traditional bibliometrics), but also Mendeley data (a new altmetrics source). Thus, the institutional impact measurement of the tool has been expanded by focusing on additional status groups besides researchers, such as students and librarians. Furthermore, the visualization of the data has been completely updated by improving the operability for the user and including new features, such as institutional profile pages.

With the consideration of Mendeley data in the new release of the tool, we focus on just one single data source that operates under the name of altmetrics – and altmetrics combine a diverse set of many data sources (National Information Standards Organization, 2016; Sugimoto et al., 2017). According to Halevi and Schimming (2018), altmetrics might "capture both public perception and overall understanding of scientific reporting". In recent years, however, altmetrics have been assessed very critically as a set of indicators that might be used in research evaluation processes (see, e.g., Rowlands, 2018). For Wilsdon et al. (2015), "the systematic use of alternative indicators as pure indicators of academic quality seems unlikely at the current time, though they have the potential to provide an alternative perspective on research dissemination, reach and 'impact' in its broadest sense" (p. 45). Robinson-Garcia et al. (2017) question the approach of "translating the citation analogy to social media … [as] the most appropriate way to capture social interactions between scholars and lay". The critique has mainly focused, however, on altmetrics data sources other than Mendeley.

It is an advantage of Mendeley data that research on altmetrics has revealed its relationship to academic impact; for many other altmetrics such as Twitter or Facebook data, it is not clear what kind of impact they measure (Tunger, Clermont, & Meier, 2018; Wilsdon et al., 2015). Another advantage of Mendeley data is its speed: since bookmarking (reading) happens before citing, Mendeley impact data are available earlier than citation data (Moed & Halevi, 2015). Haustein (2014) summarizes the advantages of Mendeley as follows: "Mendeley may be the most promising new source for evaluation purposes as it has the largest user population, the greatest coverage, highest number of readers per document, and strongest correlations between usage and citation counts". Thelwall and Kousha (2015) and Zahedi, Costas, and Wouters (2014) also see Mendeley as a promising data source for research evaluation, mentioning similar points to Haustein (2014). However, Thelwall and Kousha (2015) mention the following limitation of the data: "if the context of the informal evaluation suggests that the end users for research, or any particular target group for the evaluation, are unlikely to use Mendeley" (p. 612), Mendeley data should not be used for impact measurements.

The new release of the excellencemapping.net tool might be interesting for a broad range of people. Journalists can use the maps to obtain an overview of the institutional excellence landscape and to study the impact of publications on various status groups. Researchers might use the tool to assess their own institution in comparison to other, similar or nearby institutions. According to Petrovich (2020), science maps – such as the

excellencemapping.net tool – are especially interesting for the policy domain, since visualizations are usually suitable for non-experts to grasp complex information (e.g., on scientific performance) easily and quickly. Furthermore, the maps can be used for benchmarking and collaboration strategy: "a) Benchmarking: How is an organization performing compared to competitors? b) Collaboration strategy: Who are the potential collaborators that can complement the research mission of the organization?" (Petrovich, 2020).

Petrovich (2020) warns against possible limitations of science maps that may be also applicable to the excellencemapping.net tool: "science maps can help the decision-making, but they do not provide automatic answers. From this point of view, science maps are not different from any scientometric indicator: they provide partial representations of science whose correct interpretation should take into account many different factors … Not only science maps are error-prone … [but] their production involves several technical decisions that can deeply influence the final maps … It is pivotal that such decisions should be made transparent, and their consequences clear to the analysts and the policymakers, so that science maps do not turn into 'black boxes'". Against the backdrop of these limitations, science maps should be applied in science policy with a clear understanding of the limits.

In recent years, we have received a great deal of feedback on the excellencemapping.net tool that concerns not only the impact data of single institutions, but also the design of the tool (using the email address info@excellencemapping.net). We hope that the feedback will continue with the publication of the new release.

# References

Aduku, K. J., Thelwall, M., & Kousha, K. (2016). Do Mendeley reader counts reflect the scholarly impact of conference papers? An investigation of computer science and engineering fields. In I. Ràfols, J. Molas-Gallart, E. Castro-Martínez & R. Woolley (Eds.), *Proceedings of the 21 ST International Conference on Science and Technology Indicator*. València, Spain: Universitat Politècnica de València.

Aumüller, D., & Rahm, E. (2011). Affiliation analysis of database publications. *SIGMOD Record, 40*(1), 26-31.

Bar-Ilan, J., Halevi, G., & Milojević, S. (2019). Differences between altmetric data sources – A case study. *Journal of Altmetrics, 2*(1), 8. doi: 10.29024/joa.4.

Barthel, S., Tönnies, S., Köhncke, B., Siehndel, P., & Balke, W.-T. (2015). *What does Twitter measure? Influence of diverse user groups in altmetrics*. Paper presented at the ACM/IEEE Joint Conference on Digital Libraries (JCDL), Knoxville, TN, USA.

Blümel, C., Gauch, S., & Beng, F. (2017). Altmetrics and its intellectual predecessors: Patterns of argumentation and conceptual development. In P. Larédo (Ed.), *Proceedings of the Science, Technology, & Innovation Indicators Conference "Open indicators: Innovation, participation and actor-based STI indicators"*. Paris, France.

Bornmann, L. (2014). Do altmetrics point to the broader impact of research? An overview of benefits and disadvantages of altmetrics. *Journal of Informetrics, 8*(4), 895-903. doi: 10.1016/j.joi.2014.09.005.

Bornmann, L. (2015). Alternative metrics in scientometrics: A meta-analysis of research into three altmetrics. *Scientometrics, 103*(3), 1123-1144.

Bornmann, L., Bowman, B. F., Bauer, J., Marx, W., Schier, H., & Palzenberger, M. (2014). Bibliometric standards for evaluating research institutes in the natural sciences. In B. Cronin & C. Sugimoto (Eds.), *Beyond bibliometrics: Harnessing multidimensional indicators of scholarly impact* (pp. 201-223). Cambridge, MA, USA: MIT Press.

Bornmann, L., de Moya Anegón, F., & Leydesdorff, L. (2012). The new Excellence Indicator in the world report of the SCImago Institutions Rankings 2011. *Journal of Informetrics, 6*(2), 333-335. doi: 10.1016/j.joi.2011.11.006.

Bornmann, L., & Haunschild, R. (2016). Normalization of Mendeley reader impact on the reader- and paper-side: A comparison of the mean discipline normalized reader score (MDNRS) with the mean normalized reader score (MNRS) and bare reader counts. *Journal of Informetrics, 10*(3), 776-788.

Bornmann, L., & Haunschild, R. (2017). Measuring field-normalized impact of papers on specific societal groups: An altmetrics study based on Mendeley data. *Research Evaluation, 26*(3), 230-241. doi: 10.1093/reseval/rvx005.

Bornmann, L., Haunschild, R., & Adams, J. (2019). Do altmetrics assess societal impact in a comparable way to case studies? An empirical test of the convergent validity of altmetrics based on data from the UK research excellence framework (REF). *Journal of Informetrics, 13*(1), 325-340. doi: 10.1016/j.joi.2019.01.008.

Bornmann, L., Leydesdorff, L., & Wang, J. (2013). Which percentile-based approach should be preferred for calculating normalized citation impact values? An empirical comparison of five approaches including a newly developed citation-rank approach (P100). *Journal of Informetrics, 7*(4), 933-944. doi: 10.1016/j.joi.2013.09.003.

Bornmann, L., Mutz, R., & Daniel, H.-D. (2013). Multilevel-statistical reformulation of citation-based university rankings: The Leiden ranking 2011/2012. *Journal of the American Society for Information Science and Technology, 64*(8), 1649-1658. doi: doi.org/10.1002/asi.22857.

Bornmann, L., Mutz, R., Marx, W., Schier, H., & Daniel, H.-D. (2011). A multilevel modelling approach to investigating the predictive validity of editorial decisions: Do the editors of a high profile journal select manuscripts that are highly cited after publication? *Journal of the Royal Statistical Society Series A Statistics in Society, 174*(4), 857-879. doi: doi.org/10.1111/j.1467-985X.2011.00689.x.

Bornmann, L., Stefaner, M., de Moya Anegon, F., & Mutz, R. (2014). What is the effect of country-specific characteristics on the research performance of scientific institutions? Using multi-level statistical models to rank and map universities and research-focused institutions worldwide. *Journal of Informetrics, 8*(3), 581-593. doi: 10.1016/j.joi.2014.04.008.

Bornmann, L., Stefaner, M., de Moya Anegón, F., & Mutz, R. (2014). Ranking and mapping of universities and research-focused institutions worldwide based on highly-cited papers: A visualization of results from multi-level models. *Online Information Review, 38*(1), 43-58.

Bornmann, L., Stefaner, M., de Moya Anegón, F., & Mutz, R. (2015). Ranking and mappping of universities and research-focused institutions worldwide: The third release of excellencemapping.net. *COLLNET Journal of Scientometrics and Information Management, 9*(1), 61-68.

Bornmann, L., Stefaner, M., de Moya Anegón, F., & Mutz, R. (2016). Excellence networks in science: A Web-based application based on Bayesian multilevel logistic regression (BMLR) for the identification of institutions collaborating successfully. *Journal of Informetrics, 10*(1), 312-327.

Bornmann, L., & Williams, R. (2020). An evaluation of percentile measures of citation impact, and a proposal for making them better. *Scientometrics, 124*, 1457–1478. doi: 10.1007/s11192-020-03512-7.

Chen, C., & Song, M. (2017). *Representing scientific knowledge: The role of uncertainty*: Springer International Publishing.

Costas, R. (2017). Towards the social media studies of science: Social media metrics, present and future. *Bibliotecas. Anales de investigación, 13*(1), 1-5.

Couture-Beil, A. (2014). rjson: JSON for R. Retrieved 19 February 2021, from https://CRAN.R-project.org/package=rjson

Csomos, G. (2018). A spatial scientometric analysis of the publication output of cities worldwide. *Journal of Informetrics, 12*(2), 547-566.

Csomos, G., & Lengyel, B. (in press). Mapping the efficiency of international scientific collaboration between cities worldwide. *Journal of Information Science*.

Didegah, F., & Thelwall, M. (2018). Co-saved, co-tweeted, and co-cited networks. *Journal of the Association for Information Science and Technology, 69*(8), 959-973. doi: 10.1002/asi.24028.

Dowle, M., & Srinivasan, A. (2019). data.table: Extension of `data.frame`. Retrieved 19 February 2021, from https://CRAN.R-project.org/package=data.table

Frenken, K., Hardeman, S., & Hoekman, J. (2009). Spatial scientometrics: Towards a cumulative research program. *Journal of Informetrics, 3*(3), 222-232. doi: 10.1016/j.joi.2009.03.005.

Frenken, K., & Hoekman, J. (2014). Spatial scientometrics and scholarly impact: A review of recent studies, tools, and methods. In Y. Ding, R. Rousseau & D. Wolfram (Eds.), *Measuring scholarly impact* (pp. 127-146). Heidelberg, Germany: Springer International Publishing.

Goldstein, H. (2011). *Multilevel statistical models* (4th ed.). Chichester: John Wiley.

Gonzalez-Pereira, B., Guerrero-Bote, V. P., & Moya-Anegon, F. (2010). A new approach to the metric of journals' scientific prestige: The SJR indicator. *Journal of Informetrics, 4*(3), 379-391. doi: 10.1016/j.joi.2010.03.002.

González-Valiente, C. L., Pacheco-Mendoza, J., & Arencibia-Jorge, R. (2016). A review of altmetrics as an emerging discipline for research evaluation. *Learned Publishing, 29*(4), 229-238. doi: 10.1002/leap.1043.

Gorraiz, J., & Gumpenberger, C. (2021). PlumX Metrics (Plum Analytics) in Practice Handbook Bibliometrics (pp. 221-234). Berlin, Boston: De Gruyter Saur.

Greenland, S. (2000). Principles of multilevel modeling. *International Journal of Epidemiology, 29*(1), 158-167. doi: doi.org/10.1093/ije/29.1.158.

Grossetti, M., Eckert, D., Gingras, Y., Jégou, L., Larivière, V., & Milard, B. (2013). Cities and the geographical deconcentration of scientific activity: A multilevel analysis of publications (1987–2007). *Urban Studies, 51*(10), 2219–2234. doi: 10.1177/0042098013506047.

Gunn, W. (2013). Social signals reflect academic impact: What it means when a scholar adds a paper to Mendeley. *Information Standards Quarterly, 25*(2), 33-39.

Halevi, G., & Schimming, L. (2018). An initiative to track sentiments in altmetrics. *Journal of Altmetrics, 1*(1), 2. doi: 10.29024/joa.1.

Haunschild, R. (2021). Mendeley Handbook Bibliometrics (pp. 281-288). Berlin, Boston: De Gruyter Saur.

Haunschild, R., & Bornmann, L. (2016). Normalization of Mendeley reader counts for impact assessment. *Journal of Informetrics, 10*(1), 62-73. doi: 10.1016/j.joi.2015.11.003.

Haunschild, R., Leydesdorff, L., Bornmann, L., Hellsten, I., & Marx, W. (2019). Does the public discuss other topics on climate change than researchers? A comparison of networks based on author keywords and hashtags. *Journal of Informetrics, 13*(2), 695-707.

Haunschild, R., Stefaner, M., & Bornmann, L. (2015). *Who publishes, reads, and cites papers? An analysis of country information.* Paper presented at the Proceedings of ISSI 2015 - 15th International Society of Scientometrics and Informetrics Conference, Istanbul, Turkey.

Haustein, S. (2014). Readership metrics. In B. Cronin & C. R. Sugimoto (Eds.), *Beyond bibliometrics: Harnessing multi-dimensional indicators of performance* (pp. 327-344). Cambridge, MA, USA: MIT Press.

Haustein, S., & Larivière, V. (2014). *Mendeley as a source of readership by students and postdocs? Evaluating article usage by academic status.* Paper presented at the Proceedings of the IATUL Conferences. Paper 2.

Hazen, A. (1914). Storage to be provided in impounding reservoirs for municipal water supply. *Transactions of American Society of Civil Engineers, 77*, 1539-1640.

Hicks, D. J., Stahmer, C., & Smith, M. (2018). Impacting capabilities: A conceptual framework for the social value of research. *Frontiers in Research Metrics and Analytics, 3*(24). doi: 10.3389/frma.2018.00024.

Hox, J. J., Moerbeeck, M., & van de Schoot, R. (2017). *Multilevel analysis: Techniques and applications* (3rd ed.). New York: Taylor & Francis.

Hu, Z., Guo, F., & Hou, H. (2017). Mapping research spotlights for different regions in China. *Scientometrics, 110*(2), 779-790. doi: 10.1007/s11192-016-2175-z.

Kassab, O., Bornmann, L., & Haunschild, R. (2020). Can altmetrics reflect societal impact considerations?: Exploring the potential of altmetrics in the context of a sustainability science research center. *Quantitative Science Studies, 1*(2), 792-809. doi: 10.1162/qss_a_00032.

Konkiel, S., Madjarevic, N., & Rees, A. (2016). Altmetrics for librarians: 100+ tips, tricks, and examples. Retrieved February, 24, 2021, from http://dx.doi.org/10.6084/m9.figshare.3749838

Lang, D. T., & the CRAN team. (2018). RCurl: General network (HTTP/FTP/...) client interface for R. Retrieved 17 February 2021, from https://CRAN.R-project.org/package=RCurl

Maflahi, N., & Thelwall, M. (2015). When are readership counts as useful as citation counts? Scopus versus Mendeley for LIS journals. *Journal of the Association for Information Science and Technology*, n/a-n/a. doi: 10.1002/asi.23369.

Maflahi, N., & Thelwall, M. (2018). How quickly do publications get read? The evolution of Mendeley reader counts for new articles. *Journal of the Association for Information Science and Technology, 69*(1), 158-167. doi: 10.1002/asi.23909.

Maisonobe, M., Eckert, D., Grossetti, M., Jégou, L., & Milard, B. (2016). The world network of scientific collaborations between cities: Domestic or international dynamics? *Journal of Informetrics, 10*(4), 1025-1036. doi: 10.1016/j.joi.2016.06.002.

Maisonobe, M., Jégou, L., Yakimovich, N., & Cabanac, G. (2019). NETSCITY: A geospatial application to analyse and map world scale production and collaboration data between cities. In G. Catalano, C. Daraio, M. Gregori, H. F. Moed & G. Ruocco (Eds.), *Proceedings of the 17th international Conference on Scientometrics and Informetrics (ISSI 2019) with a Special STI Indicators Conference Track* (pp. 1195-1200). Sapienza University of Rome, Rome, Italy: ISSI.

Majeti, D., Akleman, E., Ahmed, M. E., Petersen, A. M., Uzzi, B., & Pavlidis, I. (2020). Scholar Plot: Design and evaluation of an information interface for faculty research performance. *Frontiers in Research Metrics and Analytics, 4*(6). doi: 10.3389/frma.2019.00006.

Mas-Bleda, A., & Thelwall, M. (2016). Can alternative indicators overcome language biases in citation counts? A comparison of Spanish and UK research. *Scientometrics, 109*(3), 2007-2030. doi: 10.1007/s11192-016-2118-8.

McLeish, B. (2021). Altmetric.com: A Brief History Handbook Bibliometrics (pp. 215-220). Berlin, Boston: De Gruyter Saur.

Moed, H. F. (2017). *Applied Evaluative Informetrics*. Heidelberg, Germany: Springer.

Moed, H. F., & Halevi, G. (2015). Multidimensional assessment of scholarly research impact. *Journal of the Association for Information Science and Technology, 66*(10), 1988-2002. doi: 10.1002/asi.23314.

Mohammadi, E., Thelwall, M., & Kousha, K. (2016). Can Mendeley bookmarks reflect readership? A survey of user motivations. *Journal of the Association for Information Science and Technology, 67*(5), 1198–1209.

Mutz, R., & Daniel, H.-D. (2007). Entwicklung eines Hochschul-Rankingverfahrens mittels Mixed-Rasch-Modell und Mehrebenenanalyse am Beispiel der Psychologie [Development of a ranking procedure by mixed Rasch model and multilevel analysis - psychology as an example]. *Diagnostica, 53*(1), 3–17. doi: doi.org/10.1026/0012-1924.53.1.3.

Mutz, R., & Daniel, H.-D. (2015). What is behind the curtain of the Leiden Ranking? *Journal of the Association for Information Science and Technology, 66*(9), 1950-1953. doi: doi.org/10.1002/asi.23360.

National Information Standards Organization. (2016). *Outputs of the NISO Alternative Assessment Metrics Project*. Baltimore, MD, USA: National Information Standards Organization (NISO).

Petrovich, E. (2020). Science mapping and science maps. Retrieved June 16, 2020, from https://www.isko.org/cyclo/science_mapping

Pooladian, A., & Borrego, Á. (2016). A longitudinal study of the bookmarking of library and information science literature in Mendeley. *Journal of Informetrics, 10*(4), 1135-1142. doi: 10.1016/j.joi.2016.10.003.

R Core Team. (2014). R: A Language and Environment for Statistical Computing (Version 3.1.2). Vienna, Austria: R Foundation for Statistical Computing. Retrieved from https://www.r-project.org/

Robinson-Garcia, N., Trivedi, R., Costas, R., Isset, K., Melkers, J., & Hicks, D. (2017). Tweeting about journal articles: Engagement, marketing or just gibberish? *Proceedings of the Science, Technology, & Innovation Indicators Conference "Open indicators: Innovation, participation and actor-based STI indicators"*. Paris, France.

Rodríguez-Navarro, A., & Narin, F. (2018). European paradox or delusion: Are European science and economy outdated? *Science and Public Policy, 45*(1), 14–23.

Rowlands, I. (2018). What are we measuring? Refocusing on some fundamentals in the age of desktop bibliometrics. *FEMS Microbiology Letters, 365*(8). doi: 10.1093/femsle/fny059.

Salinas, M., Giorgi , D., Ponchio, F., & Cignoni, P. (2019). A visualization tool for scholarly data. In M. Agus, M. Corsini & R. Pintus (Eds.), *STAG: Smart Tools and Applications in Graphics (2019)*.

SAS Institute Inc. (2014). *SAS/STAT 13.2 User`s Guide* Cary, NC: SAS Institute Inc.

Snijders, T., & Bosker, R. (2011). *Multilevel analysis: An introduction to basic and advanced multilevel modeling* (2nd ed.). London: SAGE Publications.

Stephens, J., Simonov, K., Xie, Y., Dong, Z., Wickham, H., Horner, J., . . . Warnes, G. R. (2018). Yaml: Methods to convert R data to YAML and back. Retrieved 17 February 2021, from https://CRAN.R-project.org/package=yaml

Sugimoto, C. R., Work, S., Larivière, V., & Haustein, S. (2017). Scholarly use of social media and altmetrics: A review of the literature. *Journal of the Association for Information Science and Technology, 68*(9), 2037-2062.

Thelwall, M. (2017). Are Mendeley reader counts useful impact indicators in all fields? *Scientometrics, 113*(3), 1721-1731. doi: 10.1007/s11192-017-2557-x.

Thelwall, M. (2018). Early Mendeley readers correlate with later citation counts. *Scientometrics, 115*(3), 1231-1240. doi: 10.1007/s11192-018-2715-9.

Thelwall, M., & Kousha, K. (2015). Web indicators for research evaluation. Part 2: Social media metrics. *Profesional De La Informacion, 24*(5), 607-620. doi: 10.3145/epi.2015.sep.09.

Thelwall, M., & Wilson, P. (2016). Mendeley readership altmetrics for medical articles: An analysis of 45 fields. *Journal of the Association for Information Science and Technology, 67*(8), 1962-1972. doi: 10.1002/asi.23501.

Tunger, D., Clermont, M., & Meier, A. (2018). Altmetrics: State of the art and a look into the future. *IntechOpen*. doi: 10.5772/intechopen.76874.

van Noorden, R. (2014). Online collaboration: Scientists and the social networks. *Nature, 512*(7513), 126–130.

Waltman, L. (2016). A review of the literature on citation impact indicators. *Journal of Informetrics, 10*(2), 365-391.

Waltman, L., Calero-Medina, C., Kosten, J., Noyons, E. C. M., Tijsen, R. J. W., van Eck, N. J., . . . Wouters, P. (2012). The Leiden ranking 2011/2012: Data collection, indicators, and interpretation. *Journal of the American Society for Information Science and Technology, 63*(12), 2419-2432. doi: doi.org/10.1002/asi.22708.

Waltman, L., Calero-Medina, C., Kosten, J., Noyons, E. C. M., Tijssen, R. J. W., van Eck, N. J., . . . Wouters, P. (2012). The Leiden Ranking 2011/2012: Data collection, indicators, and interpretation. *Journal of the American Society for Information Science and Technology, 63*(12), 2419-2432. doi: 10.1002/asi.22708.

Waltman, L., Rafols, I., van Eck, N. J., & Yegros, A. (2019). Supporting priority setting in science using research funding landscapes. Retrieved December 20, 2019, from

Waltman, L., & Schreiber, M. (2013). On the calculation of percentile-based bibliometric indicators. *Journal of the American Society for Information Science and Technology, 64*(2), 372-379.

Wickham, H. (2011). The split-apply-combine strategy for data analysis. *Journal of Statistical Software, 40*(1), 1-29. doi: 10.18637/jss.v040.i01.

Wickham, H. (2017a). Httr: Tools for working with URLs and HTTP. Retrieved 17 February 2021, from https://CRAN.R-project.org/package=httr

Wickham, H. (2017b). Tidyverse: Easily install and load the 'Tidyverse'. R package version 1.2.1. Retrieved 22 June 2020, from https://CRAN.R-project.org/package=tidyverse

Wilsdon, J., Allen, L., Belfiore, E., Campbell, P., Curry, S., Hill, S., . . . Johnson, B. (2015). *The metric tide: Report of the independent review of the role of metrics in research assessment and management*. Bristol, UK: Higher Education Funding Council for England (HEFCE).

Zahedi, Z., Costas, R., & Wouters, P. (2014). How well developed are altmetrics? A cross-disciplinary analysis of the presence of 'alternative metrics' in scientific publications. *Scientometrics, 101*(2), 1491-1513. doi: 10.1007/s11192-014-1264-0.

Zahedi, Z., & Haustein, S. (2018). On the relationships between bibliographic characteristics of scientific documents and citation and Mendeley readership counts: A large-scale analysis of Web of Science publications. *Journal of Informetrics, 12*(1), 191-202. doi: 10.1016/j.joi.2017.12.005.

Zahedi, Z., & van Eck, N. J. (2018). Exploring topics of interest of Mendeley users. *Journal of Altmetrics, 1*(1), 5. doi: 10.29024/joa.7.

https://rori.figshare.com/articles/Supporting_priority_setting_in_science_using_research_funding_landscapes/9917825

Waltman, L., & Schreiber, M. (2013). On the calculation of percentile-based bibliometric indicators. *Journal of the American Society for Information Science and Technology, 64*(2), 372-379.

Wickham, H. (2011). The split-apply-combine strategy for data analysis. *Journal of Statistical Software, 40*(1), 1-29. doi: 10.18637/jss.v040.i01.

Wickham, H. (2017a). Httr: Tools for working with URLs and HTTP. Retrieved 17 February 2021, from https://CRAN.R-project.org/package=httr

Wickham, H. (2017b). Tidyverse: Easily install and load the 'Tidyverse'. R package version 1.2.1. Retrieved 22 June 2020, from https://CRAN.R-project.org/package=tidyverse

Wilsdon, J., Allen, L., Belfiore, E., Campbell, P., Curry, S., Hill, S., . . . Johnson, B. (2015). *The metric tide: Report of the independent review of the role of metrics in research assessment and management*. Bristol, UK: Higher Education Funding Council for England (HEFCE).

Zahedi, Z., Costas, R., & Wouters, P. (2014). How well developed are altmetrics? A cross-disciplinary analysis of the presence of 'alternative metrics' in scientific publications. *Scientometrics, 101*(2), 1491-1513. doi: 10.1007/s11192-014-1264-0.

Zahedi, Z., & Haustein, S. (2018). On the relationships between bibliographic characteristics of scientific documents and citation and Mendeley readership counts: A large-scale analysis of Web of Science publications. *Journal of Informetrics, 12*(1), 191-202. doi: 10.1016/j.joi.2017.12.005.

Zahedi, Z., & van Eck, N. J. (2018). Exploring topics of interest of Mendeley users. *Journal of Altmetrics, 1*(1), 5. doi: 10.29024/joa.7.